\documentclass[12pt,preprint,color]{aastex}

\shorttitle{SMC chemical enrichment history}
\shortauthors{Carrera et al.}

\usepackage{color}
\usepackage[normalem]{ulem}

\begin{document}

\title{The Chemical Enrichment History of the Small Magellanic Cloud and Its Gradients\footnote{Based on observations collected at the European Southern Observatory, Chile, within the observing programs 074.B-0446}}


\author{Ricardo Carrera\altaffilmark{1}, Carme Gallart and Antonio Aparicio\altaffilmark{2}}
\affil{Instituto de Astrof\'{\i}sica de Canarias, Spain}
\email{rcarrera@iac.es}
\email{carme@iac.es}

\author{Edgardo Costa and Rene A. M\'endez}
\affil{Departamento de Astronom\'{\i}a, Universidad de Chile, Casilla 36--D, Santiago, Chile}


\and

\author{Noelia E. D. No\"el}
\affil{Instituto de Astrof\'{\i}sica de Canarias, Spain}
\altaffiltext{1}{Currently at Osservatorio Astronomico di Bologna, Via Ranzani 1, I-40127 Bologna, Italy; ricardo.carrera@bo.astro.it}
\altaffiltext{2}{Departamento de Astrof\'{\i}sica, Universidad de La Laguna, Spain}

\begin{abstract} 

We present stellar metallicities derived from Ca II triplet spectroscopy
in over 350 red giant branch stars in 13 fields distributed in different
positions in the SMC, ranging from $\sim$1\arcdeg\@ to $\sim$4\arcdeg\@
from its center. In the innermost fields the average metallicity is [Fe/H]
$\sim -1$. This value decreases when we move away towards outermost
regions. This is the first detection of a metallicity gradient in this
galaxy.  We show that the metallicity gradient is related to an age
gradient, in the sense that more metal-rich stars, which are also younger,
are concentrated in the central regions of the galaxy. 

\end{abstract}




\section{Introduction}

The chemical enrichment history of a galaxy is related to the origin and
distribution of nuclear species in its stars and gas. The chemical
elements are mainly produced by stars which drive the enrichment of the
interstellar medium by ejecting material containing the product of the
stellar nucleosynthesis from which the new generations of stars are
formed. In addition, gas flows also play an important role in chemical
enrichment, diluting the products of the stellar nucleosynthesis with
unenriched material from outside the galaxy, and mixing metals from one
part of the system to another (e.g. bringing metal-rich gas into
metal-poor regions). Thus, the study of chemical evolution of galaxies
involves understanding the spatial distribution and temporal evolution of
various elements by taking into account the processes of star formation,
the distribution of stars according to their masses and chemical
compositions, and the final yields of various elements and detectable
remnants of parent stars. Until recently, only the chemical enrichment
history of the solar vicinity could be studied in detail. However, the
modern multiobject spectrographers attached to the 8-10 m class telescopes
allow us to study the chemical enrichment history of the nearest Local
Group galaxies.

Because of their proximity, and the fact that they present a wide range of
ages and metallicities, the Magellanic Clouds are attractive objects to
study chemical enrichment histories. In a previous paper
\citep[][hereafter paper III]{carrera07b}, we investigated the chemical
enrichment history of the LMC. In this paper, we will focus in the study
of the SMC. There are considerable less studies of the SMC as compared
with the LMC, probably due to: (i) its irregular appearance, with complex
kinematics; (ii) its distance, located further away than the LMC; and
(iii) its depth in the line of sight, which remains a subject of
controversy.

Most of the information about the stellar populations of the SMC has been
obtained from its cluster system
\citep[e.g.][]{costahatz98,piatti01,piatti05a}. The cluster age
distribution does not show the age-gap observed in the LMC
\citep{costahatz98,mighell98}. From a sample of seven clusters older than
1 Gyr, \citet{rich00} suggest that star formation was stronger in two main
episodes, one $\sim$8$\pm$2 Gyr ago and another $\sim$2$\pm$0.5 Gyr ago.
However, there does not seem to be any age interval completely lacking of
objects \citep{rafelski05}. There is only one old cluster, NGC 121, that
is younger than most of the LMC globular clusters \citep{piatti05a}.
 However, the SMC shows a significant old field population
\citep[hereafter Paper I]{noel07}.

To our knowledge, there are only three studies in which the star formation
history (SFH) of the SMC field population was derived
\citep{dolphin01,harriszaritsky04}. From a deep color--magnitude diagram
(CMD) \citet{dolphin01} found than the star formation in a small field in 
the periphery of this galaxy was relatively constant until about 2 Gyr
ago, with no star formation occurring since then. However, this could be
biased by the fact that their field was specifically chosen because it did
not have young stars. \citet{harriszaritsky04} studied a
4\arcdeg$\times$4$\fdg$5 field in the central region of the galaxy. From
shallower CMDs they found that the star formation in the SMC has had two
main episodes: one which formed the oldest populations and lasted until
8.5 Gyr ago, and a recent one that started around 3 Gyr ago. No\"el et al
(2008, in preparation) have also obtained accurate SFHs for the fields
presented in this paper, using CMDs reaching the oldest main sequence
turnoffs with good photometric accuracy. They find that two main episodes
of star formation in all fields, one at old ages ($\simeq 10$ Gyr ago) and
another one at intermediate ages ($\simeq 5$ Gyr ago), in addition to
young star formation in the wing fields.

 Detailed determinations of chemical abundances exist only for the
youngest population of the SMC \citep[i.e.][]{hill97,venn99,hunter07}, and
its chemical enrichment history has been mainly determined from studies
from its cluster system. The cluster age--metallicity relationship (AMR)
has been obtained by \citet{piatti05a} and \citet{mighell98}, mainly from
photometric indicators. An initial chemical enrichment has been found,
followed by a period of relatively slow increase in the metal abundance.
Clusters more metal-rich than [Fe/H]$\geq$-1 are younger than 5 Gyr. Since
then, the metallicity has again increased until now. On average, the SMC
is more metal-poor than the LMC. The very recent work by \cite{idiart07},
which have obtained chemical abundances in a sample of SMC planetary
nebulae, has found a similar result, with the exception that the chemical
enrichment episode at a very early epoch is not observed.

In the present work we  focus on obtaining stellar metallicities of
individual red giant branch (RGB) stars in the field population of the SMC
from Ca II triplet (hereafter CaT) spectroscopy. These stars have been
selected in 13 fields distributed at different positions in the SMC
ranging from $\sim$1\arcdeg\@ to $\sim$4\arcdeg\@ from its center. Deep
photometry of these fields has been presented in Paper I . The procedure
followed to select the targets is explained in Section
\ref{targetselection}. The data reduction is discussed in Section
\ref{datareduction}. The radial velocities of the stars in our sample are
obtained in Section \ref{radialvelocities}. In Section \ref{cat} we
discuss the calculation of the CaT equivalent widths and the determination
of metallicities. Section \ref{agedetermination} presents the method used
to derive ages for each star by combining the information on their
metallicity and position on the CMD. The analysis of the data is presented
in Section \ref{analysis}. The metallicity distribution of each field and
the possible presence of a metallicity gradient is discussed in Section
\ref{analisismetallicitydistribution}. The derived AMRs for each field are
presented in Section \ref{agemetallicity}. The main results of this paper are discussed in Section
\ref{discursion}.

\section{Target Selection\label{targetselection}}

In the framework of a large program to obtain proper motions, deep CMDs
and stellar  metallicities in the SMC, we secured spectroscopy of stars in
13 fields spread about the galaxy body. The photometry of
these fields, presented in Paper I and  listed in Table \ref{obsfields},
were obtained with a Tektronic 2048$\times$2048 CCD detector attached to
the LCO 100\arcsec\@ telescope, which covers a field size of
8\farcm85$\times$ 8\farcm85. Following the notation described by
\citet{tinney97}, fields denoted by ``qj'' (followed by right ascension)
are centered on quasars and  were observed photometrically with the main
objective of determining the absolute proper motion of the SMC (Costa et
al., in preparation). Fields labeled ``smc'' were selected specifically to
study their stellar populations by sampling a range of galactocentric
radius at similar azimuth. The $BR$ photometry  of these fields is
described in detail in Paper I, where the distribution of stellar
populations of the SMC is discussed on the basis of CMDs. These
observations have been complemented with observations in the $I$ band in
order to allow using the reduced equivalent width-metallicity
(W'$_I$-[Fe/H]) relationship derived in \citet[hereafter Paper
II]{carrera07a} to obtain metallicities for individual RGB stars.
$I$--band observations of ``qj'' fields were obtained with the same
instrument and telescope as the BR photometry. The $I$ images of ``smc''
fields were obtained with FORS2 at the VLT in image mode and were also
used for spectroscopic mask configuration of MXU@FORS2. Field qj0111 was
observed with both telescopes in order to compare the photometric
calibrations. The magnitudes obtained with each calibration differ by
about $\sim$0.1 magnitudes. This relatively large difference is mainly
explained by the poor photometric calibration of the FORS2 images, which
were not taken for photometric purposes. In any case, this error means a
metallicity uncertainty of only $\sim$0.02 dex, a value smaller than the
metallicity uncertainty itself ($\sim$0.1 dex).

\begin{figure}
\plotone{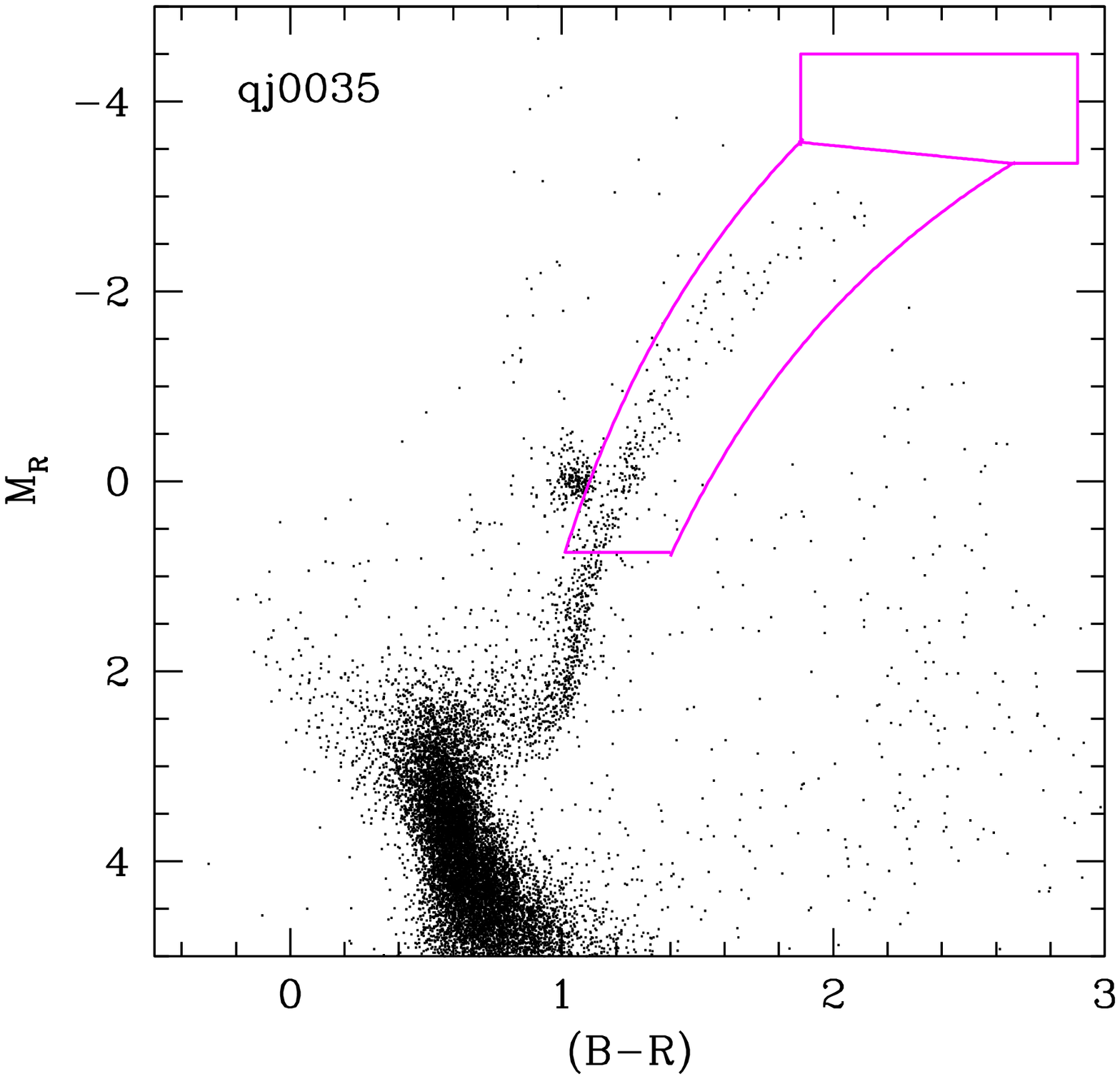}
\caption{Color--magnitude diagram of the field qj0035 showing the RGB windows used to select the candidates to be observed spectroscopically.\label{dcmbox}}
\end{figure}

In each field we selected stars in two windows of the CMD, which are
plotted in Figure \ref{dcmbox}. In each region the stars were ordered from
the brightest to the faintest ones, regardless of their color. The
resulting star list was used as input for the mask configuration task of
the instrument. Stars in the box below the RGB tip were given higher
priority, and only objects above the tip were observed when it was
impossible to put the slit on a star of the lower region. 

\section{Observations and Data Reduction\label{datareduction}}

The spectroscopic observations were carried on in service mode with the
VLT ANTU telescope, at Paranal Observatory (Chile), through program
074.B-0446. We used FORS2 in MXU multiobject mode with  grism 1028z+29 and
filter OG590+32 in order to eliminate residual orders. Slits with a length
of 8\arcsec\@ and a width of 1\arcsec\@ were selected. With this setup, we
were able to observe about 40 objects in each field (the final number
depends on the spatial distribution of the selected stars). The slit
length was selected with the purpose of shifting the stars along  it in
order to acquire two exposures of each field without  superposition of the
stellar spectra in each. This particular procedure allowed to extract the
spectra in the same way as Paper II. 

After bias subtraction, each image was flat-field corrected. Then,  as
each star is in a different position in the two images of the same field,
we subtracted one from the other, obtaining a positive and a negative
spectrum of the same star. With this procedure the sky is subtracted in
the same  pixel in which the star has been observed, thus, minimizing the
effects of pixel-to-pixel sensitivity variations. Sky residuals due to
temporal variation of the sky brightness were eliminated in the following
step, in which the spectrum is extracted in the usual way and the
remaining sky background is subtracted using the information on both sides
of the stellar spectrum. In the next step, the spectra were wavelength
calibrated and added to obtain the final spectrum. Finally, each spectrum
was  normalized by fitting a polynomial, excluding the strongest lines
(such as the CaT lines). There is an uncertainty in the wavelength
calibration because the arcs used for this purpose were not taken at the
same time and with the same telescope pointing as the object. The effects
of this on the wavelength calibration have been discussed by
\citet{gallart01}. Since we are not interested in obtaining precise radial
velocities, this problem will not significantly affect our results.

In total, we observed 386 stars in the 13 SMC fields. Their magnitudes and
CaT equivalent widths are listed in Table \ref{starobs}

\section{Radial Velocities\label{radialvelocities}}

The radial velocity of each star was calculated in order to reject SMC
non-members. We used the IRAF \textsl{fxcor} task, which performs the
cross-correlation between the target and template spectra of stars of
known radial velocity \citep{td79}. As templates we selected nine stars in
the clusters NGC 104, NGC 2682, NGC 288 and NGC 7078 which were observed
within the same program as the SMC fields and presented in Paper II. The
velocities were corrected to the heliocentric reference frame within
\textsl{fxcor}. The final radial velocity for each SMC star is the average
of the velocities obtained from each template, weighted by the width of
the corresponding correlation peaks. The resulting velocities can be
affected by the fact that the stars might have not been positioned exactly
in the center of the slit. The importance of this is described in detail
in Paper II. However, in the case of the SMC, the stars were observed at a
relatively large air mass ($\geq$1.6) and with a seeing near to or larger
than 1\arcsec. Since the slit width was 1\arcsec, the effect of the
incorrect centering of the star on the slit was of little importance. In
fact, when we tried to characterize this effect in the way described in
Paper II, we found that its contribution to the resulting velocity is
smaller than the uncertainty due to the wavelength calibration. For this
reason, we did not take this effect into account in the final radial
velocity. We considered as SMC members those stars with radial velocities
in the range 50$\leq$V$_r\leq$250 km s$^{-1}$ \citep{harriszaritsky06}.

\section{CaT Equivalent Widths and Metallicity determination\label{cat}}

The metallicity of the RGB stars is obtained following the procedure
described in Paper II. The equivalent width is the area of the line
normalized to the local continuum within a line bandpass. The continuum is
calculated from a linear fit to the mean value of the corresponding
bandpasses. The line and continuum bandpasses used in this work are listed
in Table \ref{bandastable}. The line flux is calculated from the fit of
its profile using a Gaussian plus a Lorentzian function. As discussed in
Paper II, this function provides a better fit to the core and the wings of
the strongest lines than other functions previously used. The CaT index,
denoted as $\Sigma Ca$, is defined as the sum of the equivalent widths of
the three CaT lines. The $\Sigma Ca$ and their uncertainties for each star
observed are given in Table \ref{starobs}, together with their magnitudes
and radial velocities. Two calibrations of the CaT as metallicity
indicator were obtained in Paper II based on $I$ and $V$ magnitudes. In
this case, only $I$ magnitudes are available for the SMC stars. The
reduced equivalent width, W'$_I$, for each star has been calculated using
the slope obtained in Paper II for the calibration clusters in the
M$_I$-$\Sigma Ca$ plane ($\beta_I$=-0.611 \AA mag$^{-1}$). To obtain the
absolute magnitudes we assumed a distance modulus of (m-M)$_0$=18.9
\cite[see][]{vandenbergh99} and reddenings listed in the last column of Table
\ref{obsfields} (see Paper I for details). Also, in Paper II three
different metallicity scales were used as reference. In this case, we only
used the relationships obtained on the \citet[hereafter CG97]{cg97}
metallicity scale, because it is the only one that uses homogeneous
high-resolution metallicities of open and globular clusters, and because
the metallicities of the LMC stars in Paper III were also obtained in this
way.

In brief, the metallicity for each star is given by:

\begin{equation}\label{metaleq}
[Fe/H]_{CG97}=-2.95+0.38\Sigma Ca+0.23M_I
\end{equation}

\begin{figure}
\plotone{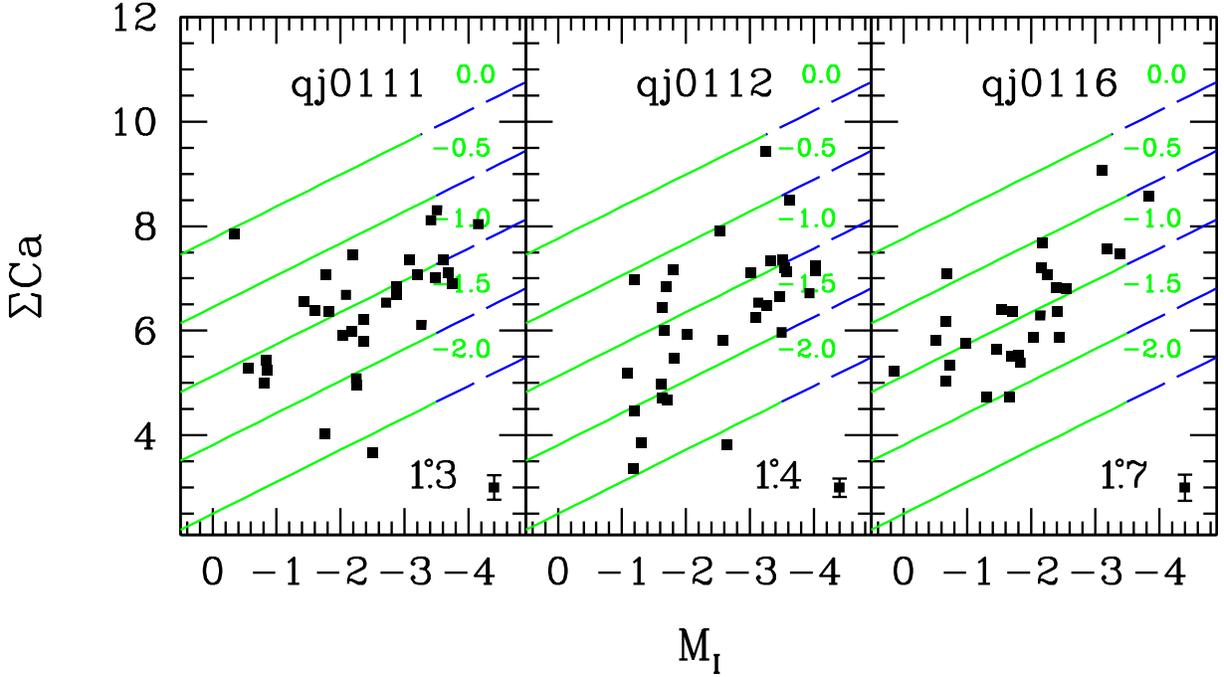}
\caption{Position in the M$_I$--$\Sigma$Ca plane of observed SMC stars in the eastern fields. Only stars with confirmed membership from their radial velocity are represented. The typical $\Sigma Ca$ error is shown in the bottom right corner of each panel.
Isometallicity lines, obtained from the relationship based on M$_I$ presented in Paper II, have been plotted
for reference. The
solid part of the line is the magnitude interval covered by the cluster stars used for the calibration
(see Paper II). The dashed part is the region in which the calibration is
extrapolated. Distances from the SMC optical center are given in the bottom right corner. The innermost field is on the left and the outermost one is on the right.\label{misigmacaeste}}
\end{figure}

\begin{figure}
\plotone{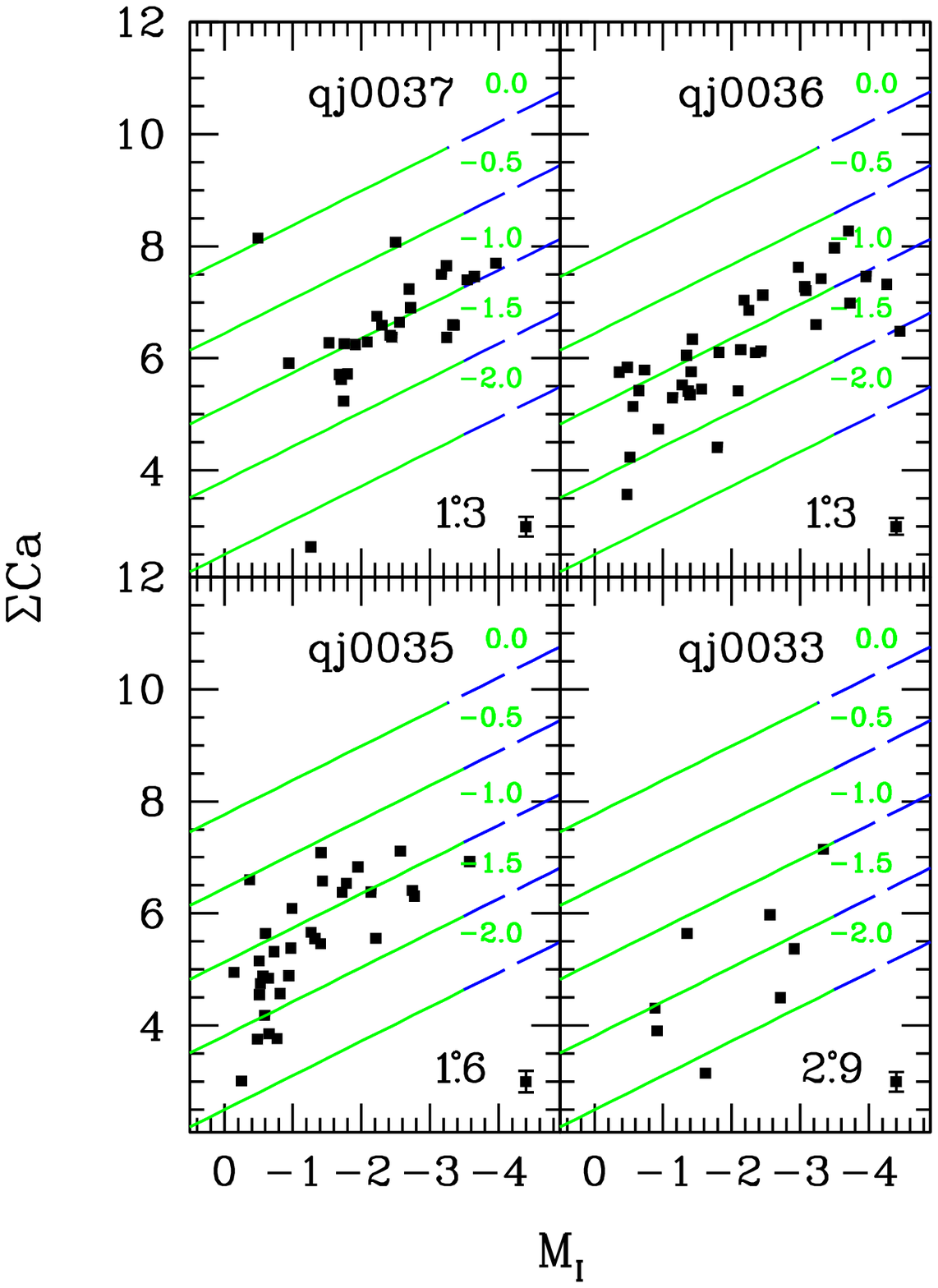}
\caption{The same as Figure \ref{misigmacaeste}, for the Western fields. They are ordered from the innermost field (\textsl{top
left}), to the outermost one (\textsl{bottom right}).\label{misigmacaoeste}}
\end{figure}

\begin{figure}
\plotone{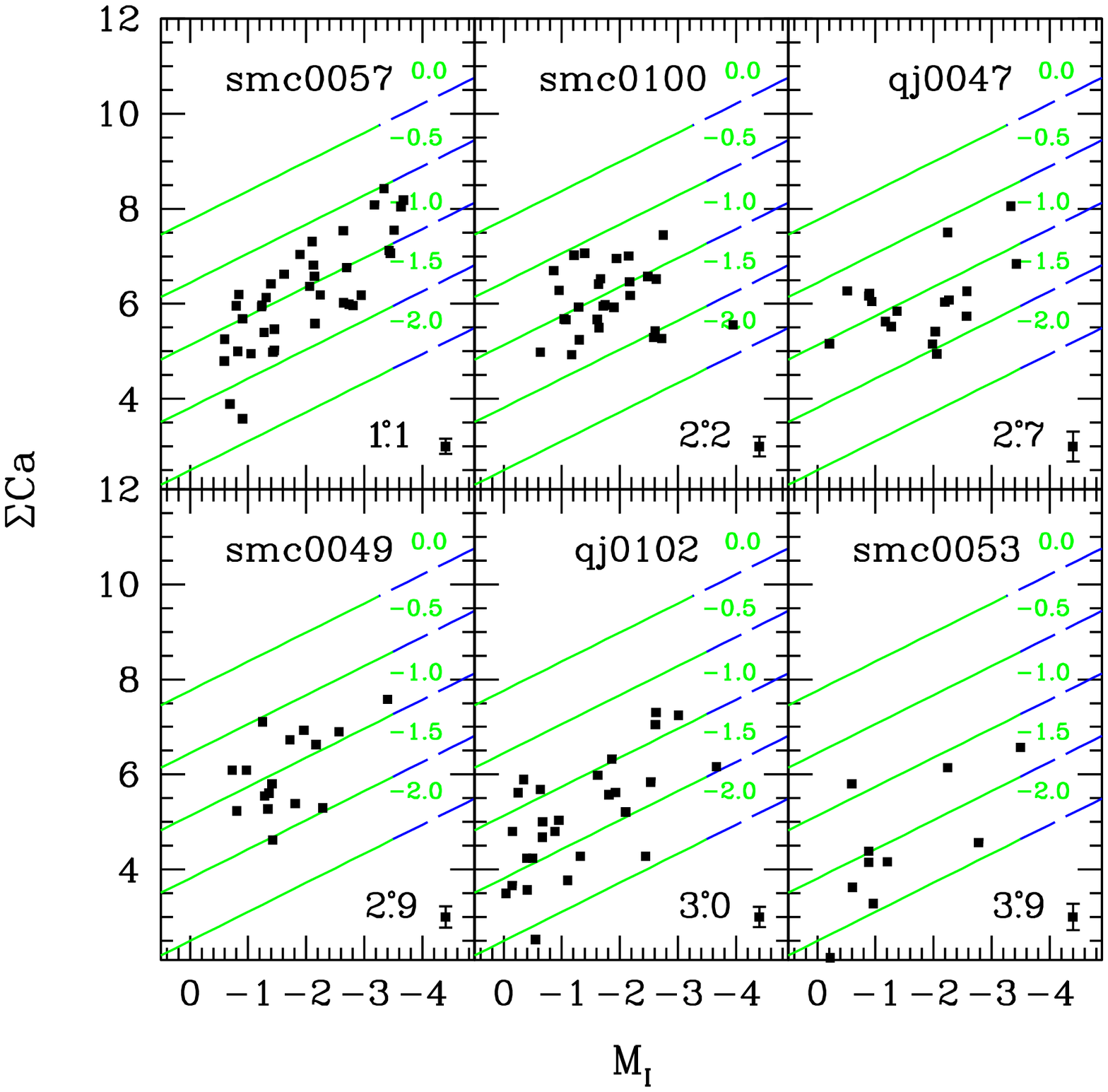}
\caption{The same as Figure \ref{misigmacaeste}, for the Southern fields. They are ordered from the innermost field (\textsl{top
left}), to the outermost one (\textsl{bottom right}).\label{misigmacasur}}
\end{figure}

In Figures \ref{misigmacaeste}, \ref{misigmacaoeste} and
\ref{misigmacasur} the position of SMC stars (radial velocity members) in
the M$_I$-$\Sigma$Ca plane for our eastern, western, and southern fields,
are respectively shown.  Solid lines indicate the magnitude range of the
observed cluster stars used in Paper II to obtain the above relationship.
Dashed lines show the region where the relationship is extrapolated.

\begin{figure}
\plotone{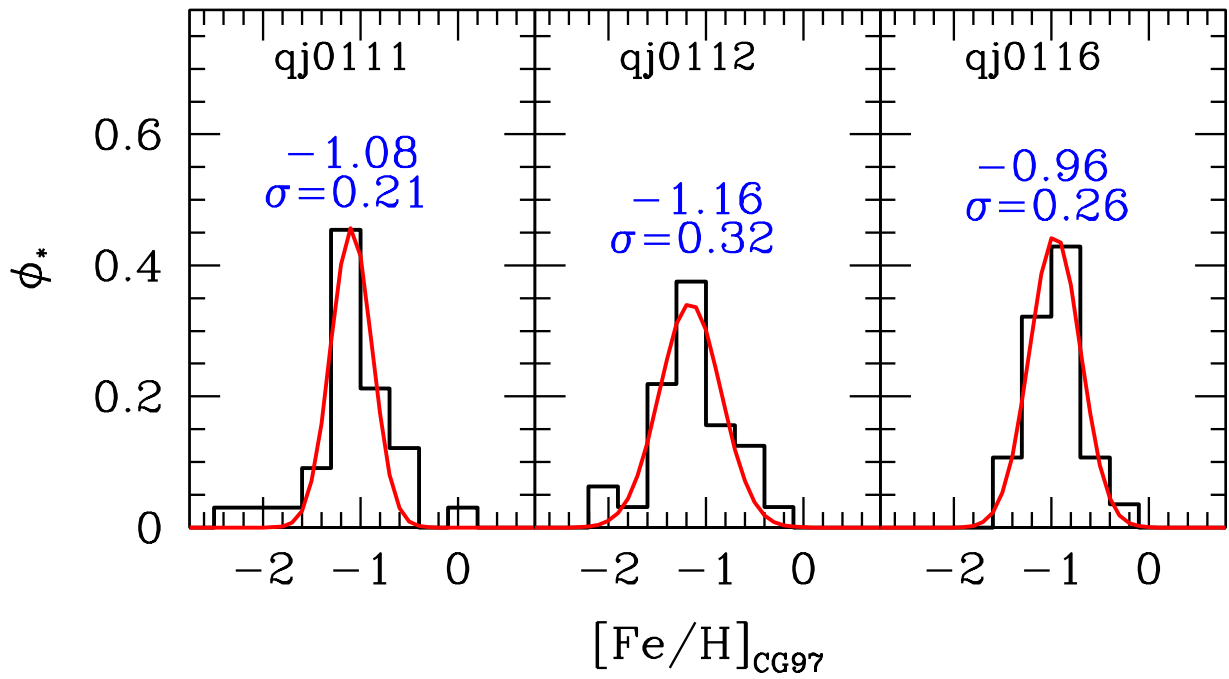}
\caption{Metallicity distributions for the eastern fields in our sample. A 
Gaussian has been fitted to each distribution in order to obtain its
mean and dispersion. The values obtained are shown in each panel.\label{metaldisteste}}
\end{figure}

\begin{figure}
\plotone{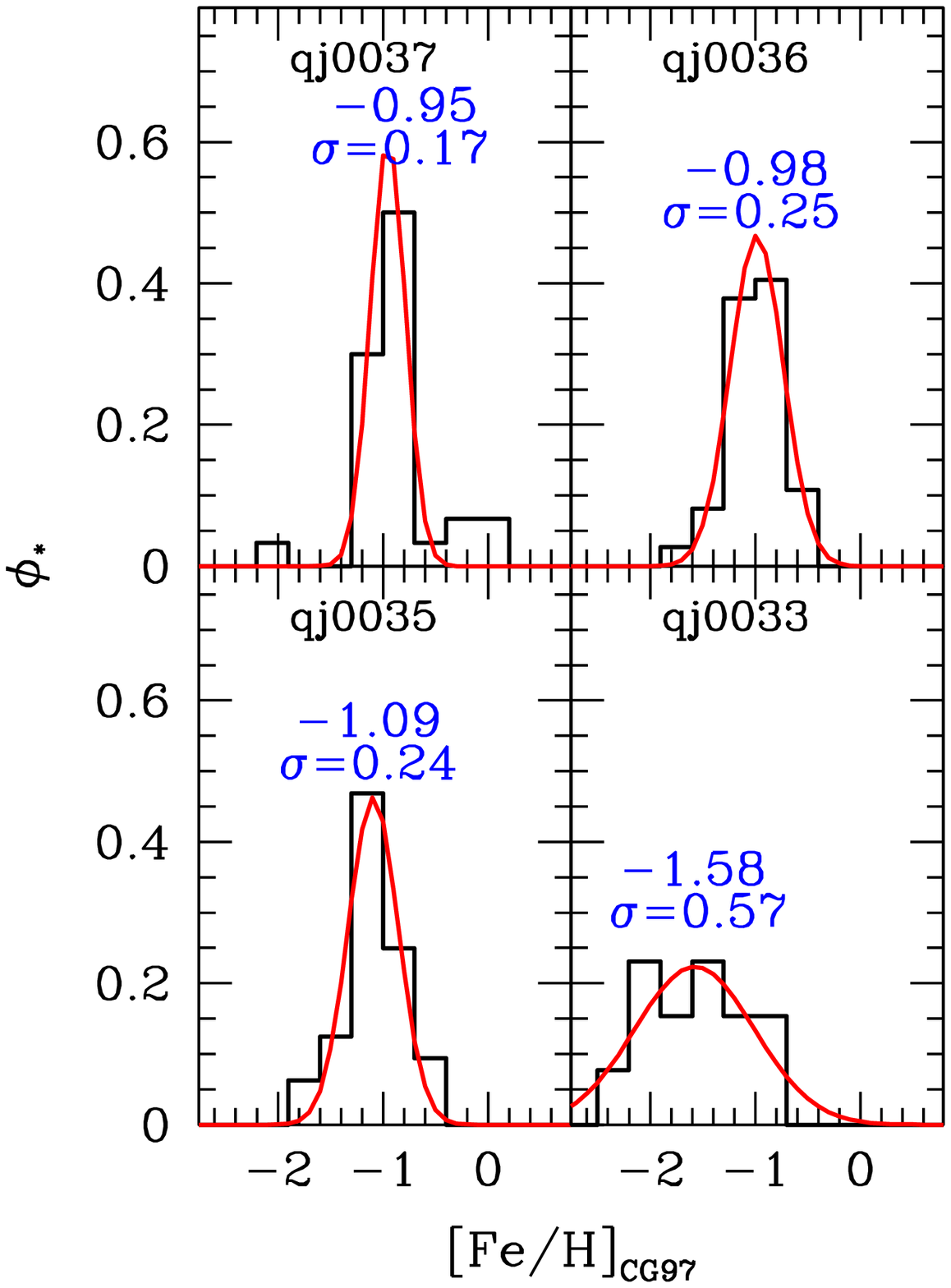}
\caption{The same as Figure \ref{metaldisteste} for the western fields.\label{metaldistoeste}}
\end{figure}

\begin{figure}
\plotone{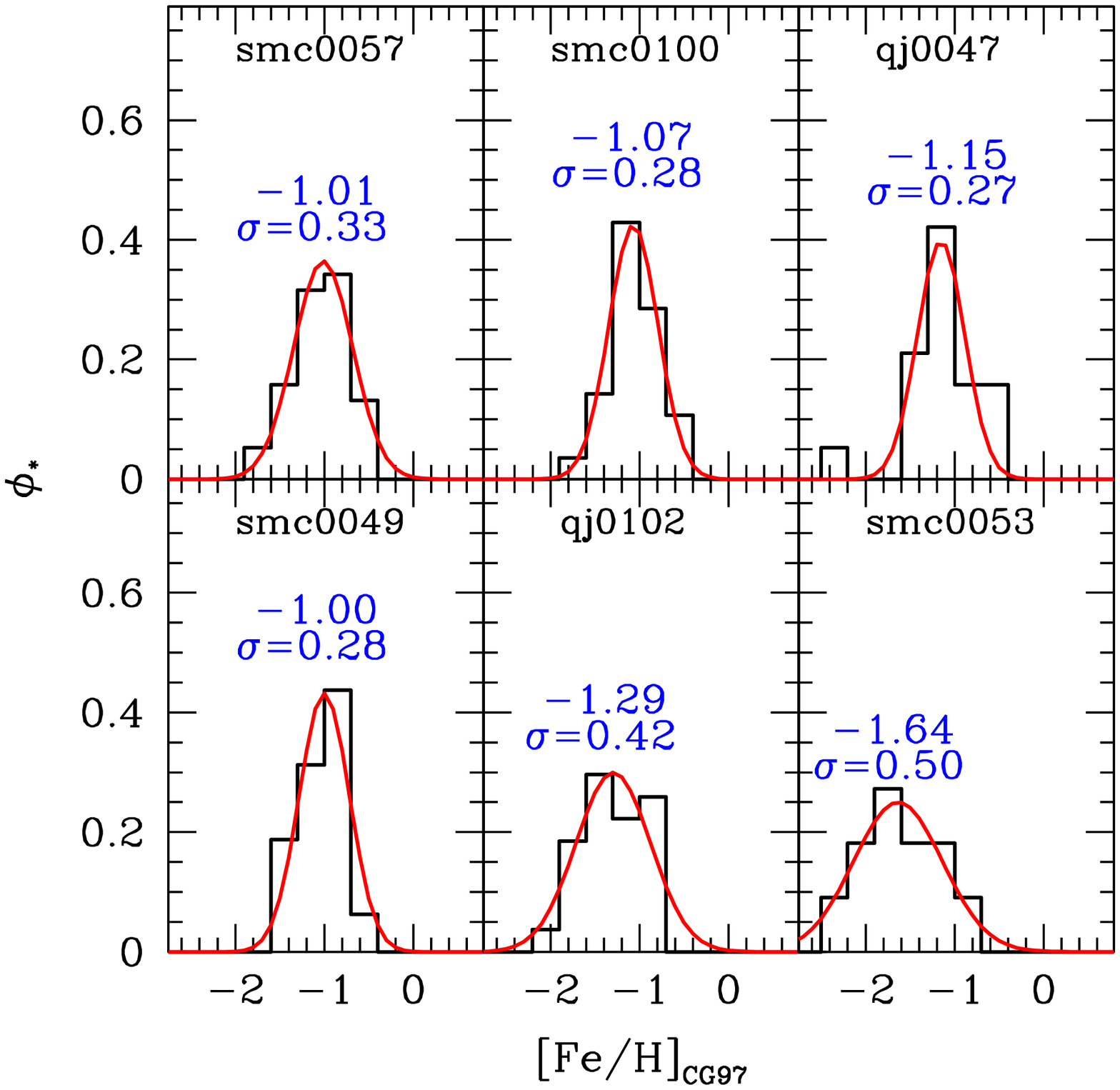}
\caption{The same as Figure \ref{metaldisteste} for the southern fields.\label{metaldistsur}}
\end{figure}

The metallicity distribution of each field is shown in Figures
\ref{metaldisteste}, \ref{metaldistoeste} and \ref{metaldistsur} and will
be discussed in  Section \ref{analisismetallicitydistribution}.

\section{Determination of stellar ages\label{agedetermination}}

The position of the RGB on the CMD suffers from age--metallicity
degeneracy. However, when the metallicity is obtained in an alternative
way, as in this case from spectroscopy, this age--metallicity degeneracy
can be broken, and ages can be derived from the position of the stars in
the CMD. In paper III, a polynomial relationship was computed to derive
stellar ages from their metallicities and positions in the CMD. For that
purpose, synthetic CMDs computed with IAC-star \citep{aparicio04} with the
overshooting BaSTI \citep{pie04} and Padova
\citep{girardi02} stellar stellar evolution models as input were used.
As it is explained in Paper III, differences between both
models in the resulting ages are negligible for our purpose, since we are
not interested in an accurate determination of ages. For simplicity, as in
Paper III, we used only the relationship obtained from the BaSTI stellar
models. The aforementioned relation was obtained for $(V-I)$ and $M_V$.
Since $V$ magnitudes are unavailable for our sample of SMC stars, we
computed a new relationship for $B$, $R$, and $I$ magnitudes. In this
case, we selected $(B-I)$ instead of $(B-R)$ because the former is much
more sensitive to small changes in the stellar metallicity and age. 

We followed the same procedure as in Paper III. First, we used the same
synthetic CMD used in Paper III  \footnote{During the referreing process
of this paper, the authors of the BaSTI models discovered a problem in the
calculation of models in the mass range 1.1--2.5 M$_\odot$ ($\simeq$1--4
Gyr), (see http://albione.oa-teramo.inaf.it/). Even though the differences
between the new and old models were unlikely to affect in any substantial
way our results, we recalculated equation \ref{rela} using a synthetic CMD
computed with the updated model set. All the results shown are derive from
this new relationship. We have verified that the differences are indeed
minor, and that, therefore, the results on the LMC on paper III, obtained
using the faulty models, can be trusted nevertheless.}, which was computed
with a constant star formation rate (SFR) between 0 and 13 Gyr and with a
chemical law such that any star can have any metallicity between $-2.3$ to
+0.5 dex. In this CMD we only selected stars in the same region below the
tip of the RGB in which the observed ones were mainly chosen. We did not
consider the brightest AGB stars due to the uncertainty of their
parameters as predicted by stellar evolution models. Following the same
statistical procedure  described in Paper III, we  checked which
polynomial combinations of magnitude $M_I$, color $(B-I)$ and metallicity
$[Fe/H]$ best represent the age of the stars in this synthetic CMD. In
order to minimize the $\sigma$ and to improve the correlation coefficient,
different linear, quadratic and cubic terms of each observed magnitude
have been added. We have checked whether $M_I$ or $M_B$ magnitudes improve
the relationship. Similar results are obtained for both magnitudes, so we
choose the first because metallicities were calculated from it. The final
polynomial form adopted is:

\begin{equation}\label{rela}
log(age)=a+b(B-I)+cM_I+d[Fe/H]+f(B-I)^2+g[Fe/H]^2
\end{equation}

The best fit coefficients are listed in Table \ref{coefedad}.

\begin{figure}
\plotone{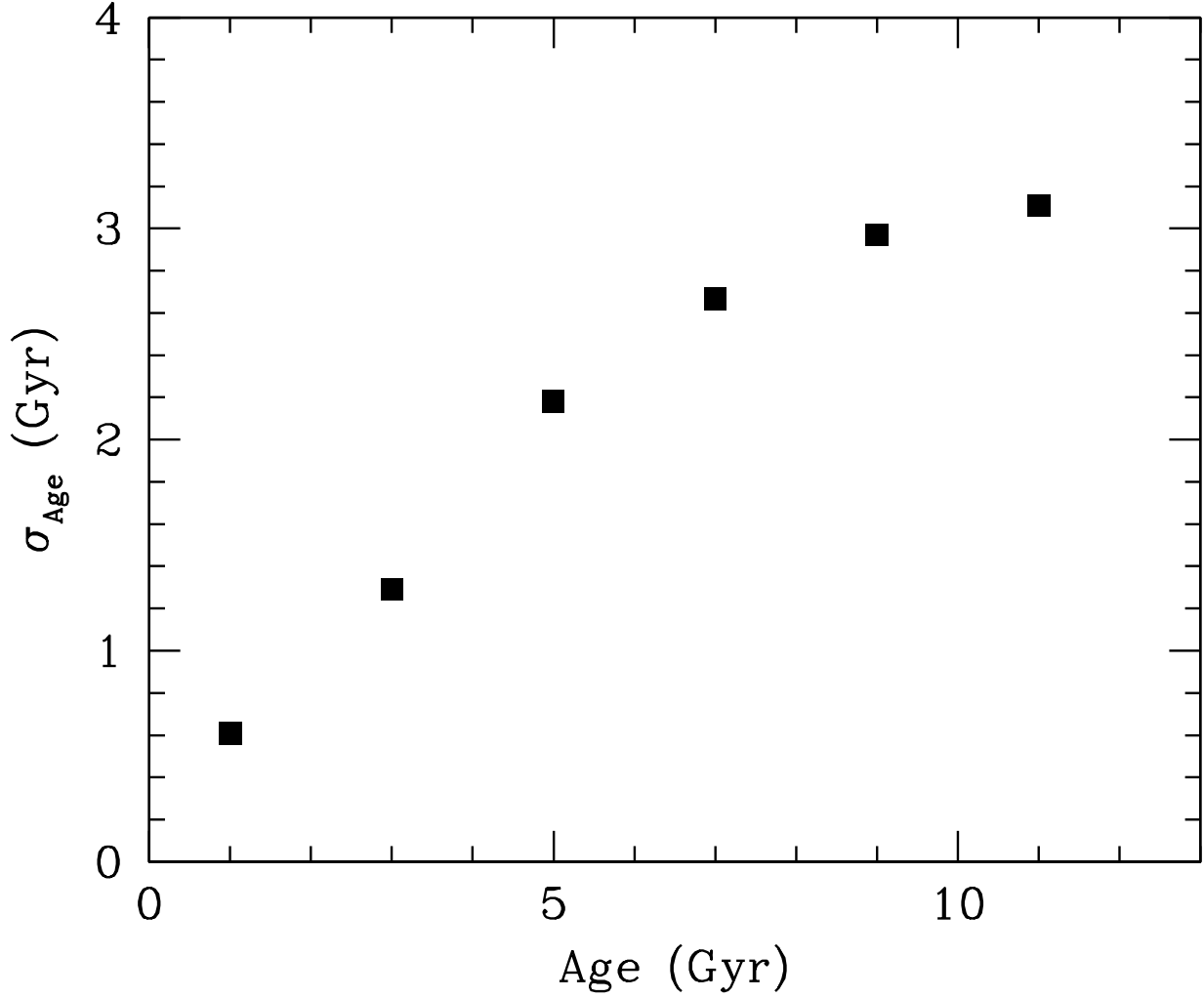}
\caption{Age uncertainty as a function of age as calculated through a Monte Carlo test and using equation \ref{rela}. 
See text for details.\label{caledad}}
\end{figure}

In order to estimate the age uncertainty when this relationship is used to
compute stellar ages, we performed a Monte Carlo test as in paper III. The
goal is to check how the obtained ages change when the input parameters
are modified.  The test consists
in computing, for each synthetic star, several age values for stochastically varying $[Fe/H]$, $(B-I)$ and $M_{I}$ according  to a gaussian probability
distribution of the corresponding $\sigma$ ($\sigma_{[Fe/H]}\sim$0.15 dex; $\sigma_{(B-I)}\sim$0.001 and
$\sigma_{M_I}\sim$ 0.001). The $\sigma$ value of the obtained ages provide an estimation of the age error when
Equation \ref{rela} is used. The obtained values for the considered age intervals are shown in
Figure \ref{caledad}. The age uncertainty increases for older ages.

\placefigure{caledad}

\begin{figure}
\plotone{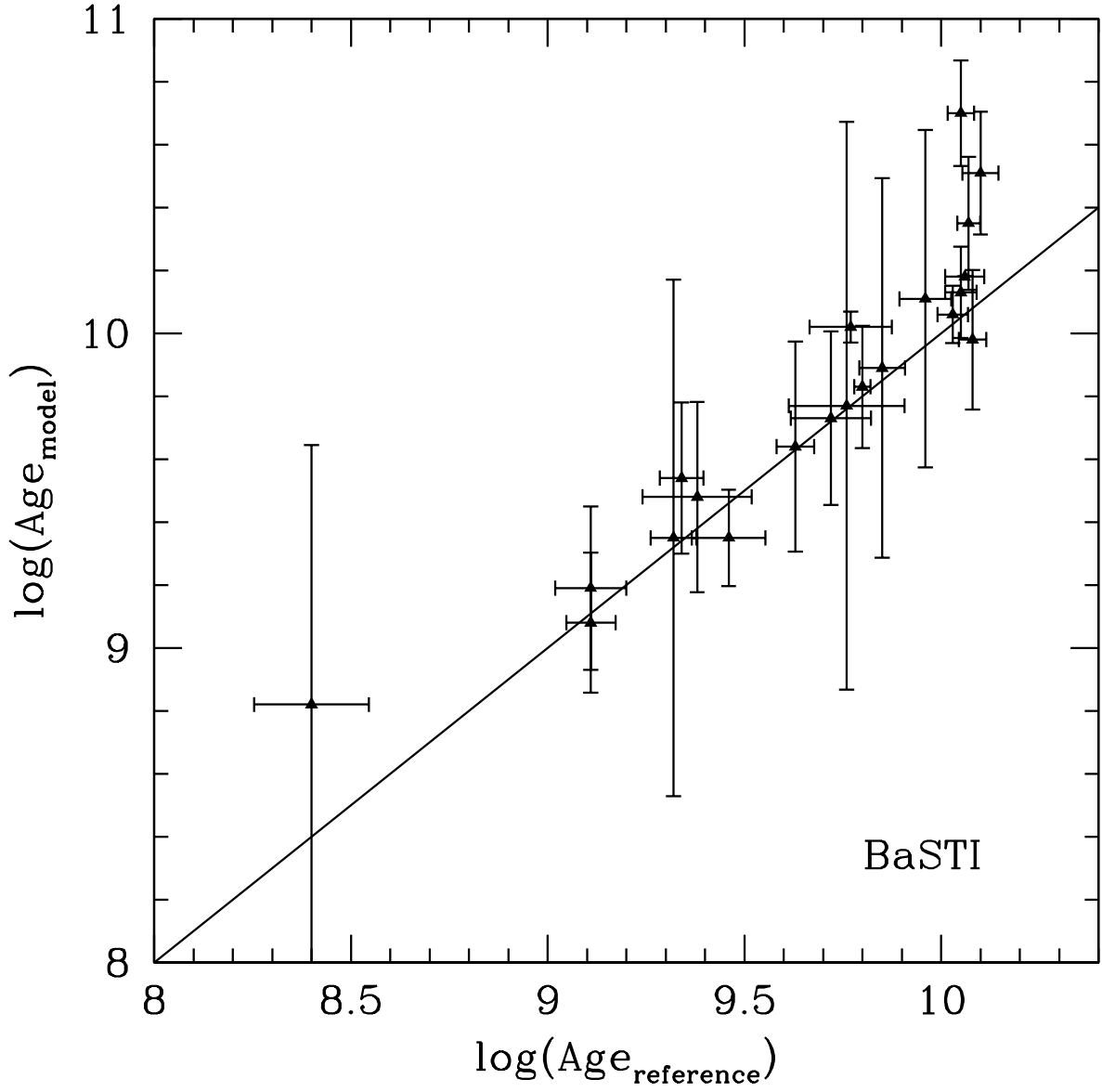}
\caption{Ages derived from equation \ref{rela} for the cluster sample presented in Paper II, plotted against the reference values. The solid line corresponds to the one-to-one relation. 
\label{cluster}}
\end{figure}

It is also necessary to check how well equation \ref{rela} reproduces the
age of a real stellar system. Following the same steps as in Paper III, we
choose the cluster stars used in Paper II for the calibration of the CaT
as metallicity indicator. For those stars we knew their $(B-I)$ color and
$M_I$  magnitudes, so we can compute their metallicity in the same way as
was done for the SMC stars (see Section \ref{cat}). We then used these
observational magnitudes as input for equation \ref{rela}, obtaining an
age for each cluster star. The cluster age was computed as the mean of the
ages of its member stars. In Figure \ref{cluster}, the age computed for
each cluster has been plotted versus its reference value. As in Paper
III,  ages younger than 10 Gyr are well recovered. However, the
relationship saturates for ages larger than 10 Gyr.

\placefigure{cluster}

\section{Analysis\label{analysis}}

\subsection{Metallicity distribution\label{analisismetallicitydistribution}}

Metallicity distributions are shown in Figures \ref{metaldisteste},
\ref{metaldistoeste} and \ref{metaldistsur} for eastern, western and
southern SMC regions respectively. We have fitted a Gaussian to obtain the
mean metallicity and metallicity dispersion of each of them. These values
are listed in columns 3 and 4 of Table \ref{smcmetalicidades}. The fields
are ordered by its distance to the center, which is shown in column 2.
Fields in different regions are indicated by different font types: eastern
fields in normal, western fields in boldface and southern fields in
italics. Mean metallicities are very close to [Fe/H]$\sim$-1 in all fields
within r$\lesssim$2$\fdg$5 from the SMC center. A similar value is
observed for the southern fields up to r$\lesssim$3\arcdeg\@ (qj0047 and
smc0049). Note that the SMC isopleths of intermediate--age and old stars
are elongated in the NE-SW direction \citep[PA=45\arcdeg;][]{cioni00} and
that a radius of 3\arcdeg\@ in the southern direction correspond to
approximately the same isopleth at radius 2$\fdg$5 in the eastern and
western directions. For the outermost fields, qj0033 in the West, and
qj0102 and qj0053 in the South, the mean value is clearly more metal-poor
than in the others.

The fact that the mean metallicity decreases when we move away from the
center implies that there is a metallicity gradient in the SMC. This
gradient is more clear when we compute the percentage of stars in
different metallicity bins, values also listed in Table
\ref{smcmetalicidades} (columns 5 and 6). For the western and southern
fields, the percentage of stars more metal-poor than [Fe/H] $=-1$
increases when we move away from the center. This is not observed in the
eastern fields because they are almost at the same distance. This is the
first time that a metallicity gradient has been reported in SMC stellar
populations. The detection of this gradient has been possible because we
have covered a large radius range, up to 4\arcdeg\@ from the SMC center.

\subsection{Age--metallicity relationships\label{agemetallicity}}

From a purely phenomenological point of
view, there are two main ways to account for the mean metallicity gradient
found in the previous section. One possibility is that chemical enrichment
has proceeded more slowly towards the SMC periphery, in such a way that
coeval stars would be more metal-poor when we move away from the center.
This seems to be the case of spiral galaxies, like the Milky Way, where the
observed abundance gradients may be explained by radial variations of the
relation between the SFR and the amount of infalling gas
\citep[e.g.][]{prantzos00,chiappini01}. The situation would be complicated
in dwarf galaxies by the probable existence of galactic winds originated
in supernova explosions \citep[e.g.][]{romano06} which are able to remove
large amount of metals from the interstellar medium. An alternative
scenario is that the stellar AMR (i.e. the law of chemical enrichment as a
function of time)  has been the same everywhere in the SMC. As a result,
the average metallicity of coeval stars would be the same in all fields
and the metallicity gradient would be related to an age gradient.
A mixture of both scenarios is also possible.

\begin{figure}
\plotone{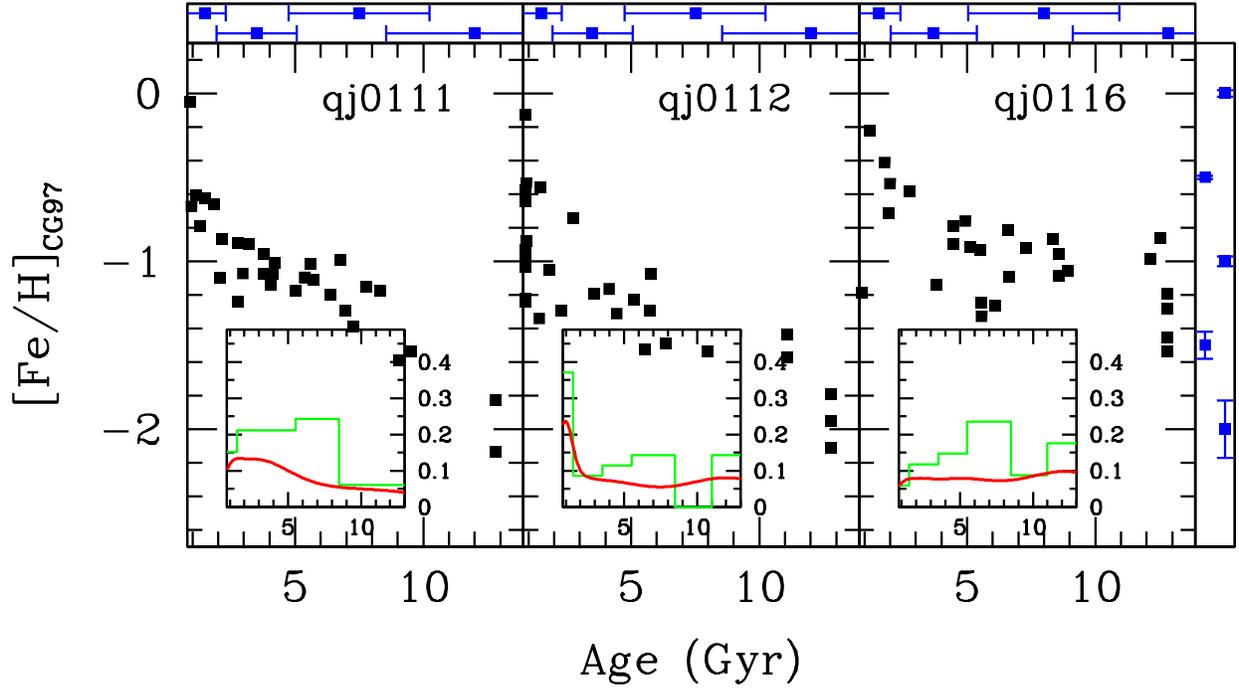}
\caption{Age--metallicity relationships for the eastern SMC fields in our 
sample. Inset panels show the age distribution computed taking 
(\textsl{solid line}) and not taking (\textsl{histogram}) into account 
the age determination uncertainties. Top panels show the age error in each 
age interval (see Figure \ref{caledad}). Right panel shows the metallicity
uncertainty in each metallicity bin. 
\label{amreste}}
\end{figure}

\begin{figure}
\plotone{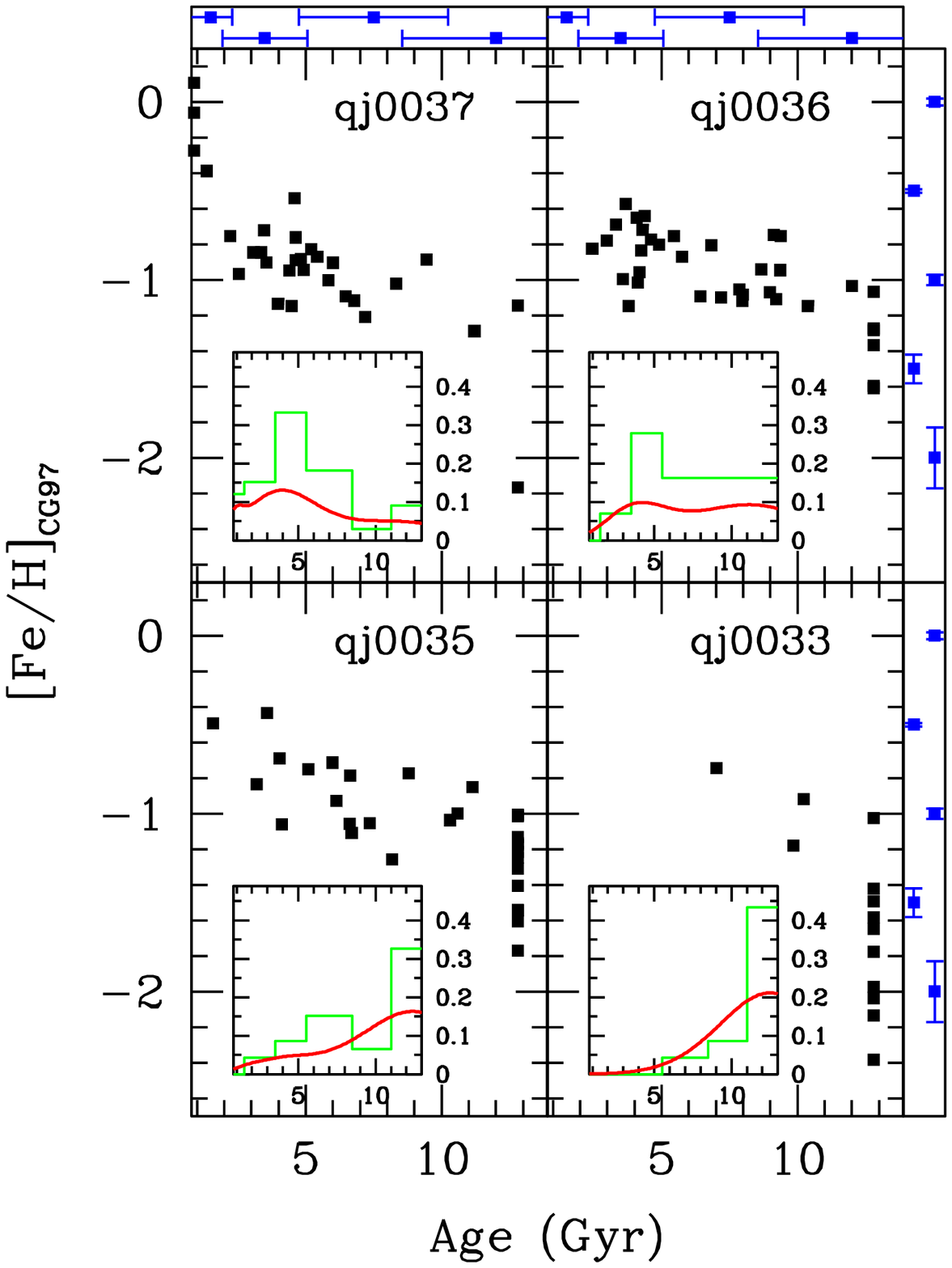}
\caption{The same as Figure \ref{amreste}, for the western fields.
\label{amroeste}}
\end{figure}

\begin{figure}
\plotone{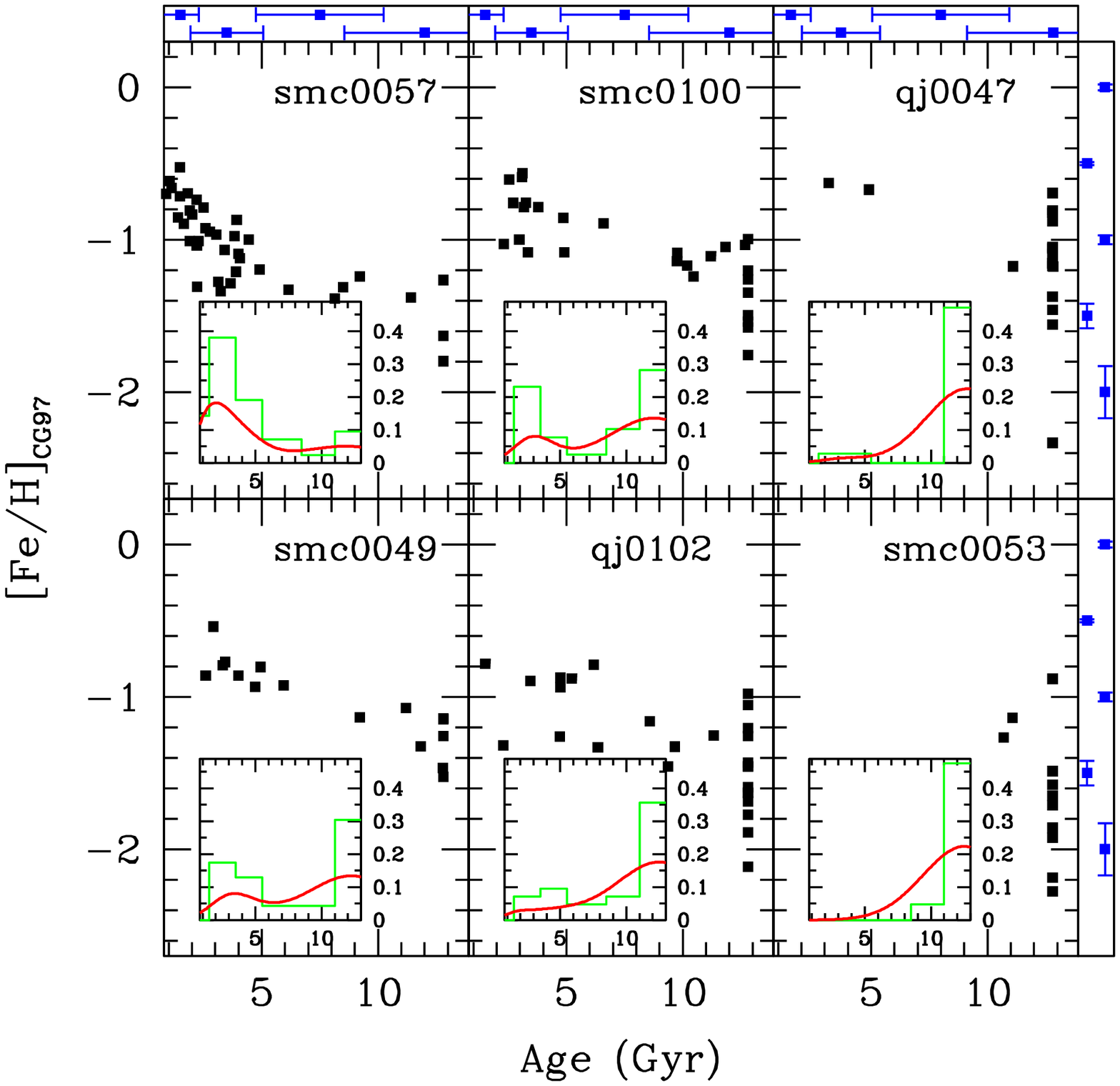}
\caption{The same as Figure \ref{amreste}, for the southern fields.
\label{amrsur}}
\end{figure}

To investigate the nature of the gradient, we have therefore calculated
the AMR for each field. They are plotted in Figures \ref{amreste},
\ref{amroeste}, and \ref{amrsur} for fields situated to the East, West and
South respectively. Note that the uncertainty in age is much larger than
in metallicity. However, as we are interested in  the global behavior and
not in obtaining individual stellar ages, the age determinations are still
valid. Since the procedure to obtain the age saturates for values older
than 10 Gyr, we can only be confident that the oldest stars have an age
$\geq$ 10 Gyr. For these we assume an age of 12.9 Gyr, which is the age of
the oldest cluster in the Milky Way \citep[NGC 6426,][]{sw02}. Regarding
the youngest stars, in the region of the CMD where we selected the
observed stars, and according to the stellar evolution models, we do not
expect to find stars younger than $\sim$0.8 Gyr. However, equation
\ref{rela} can formally compute ages younger than this value. As the age
determination uncertainty for these young stars  is $\sim$1 Gyr, in order
to avoid this contradiction we assign them an age of 0.8 Gyr. Inset panels
show the age distribution of the observed stars, with and without taking
into account the age uncertainty (\textsl{solid line and histogram},
respectively). To obtain the first one, we assumed that the age of each
star is represented by a Gaussian probability distribution on the age
axis, with a mean value equal to the age calculated for this given star,
and $\sigma$ equal to the age uncertainty. The area of each of these
distributions is unity. In the case of stars near the edges, the wings of
the distribution may extend further than the limits. We have proceeded in
the same way as described in Paper III, cutting off the wings outside the
assumed limits (0.8 and 12.9 Gyr) and rescaling the rest of the
distribution so that the area remains unity.

\begin{figure}
\epsscale{1}
\plotone{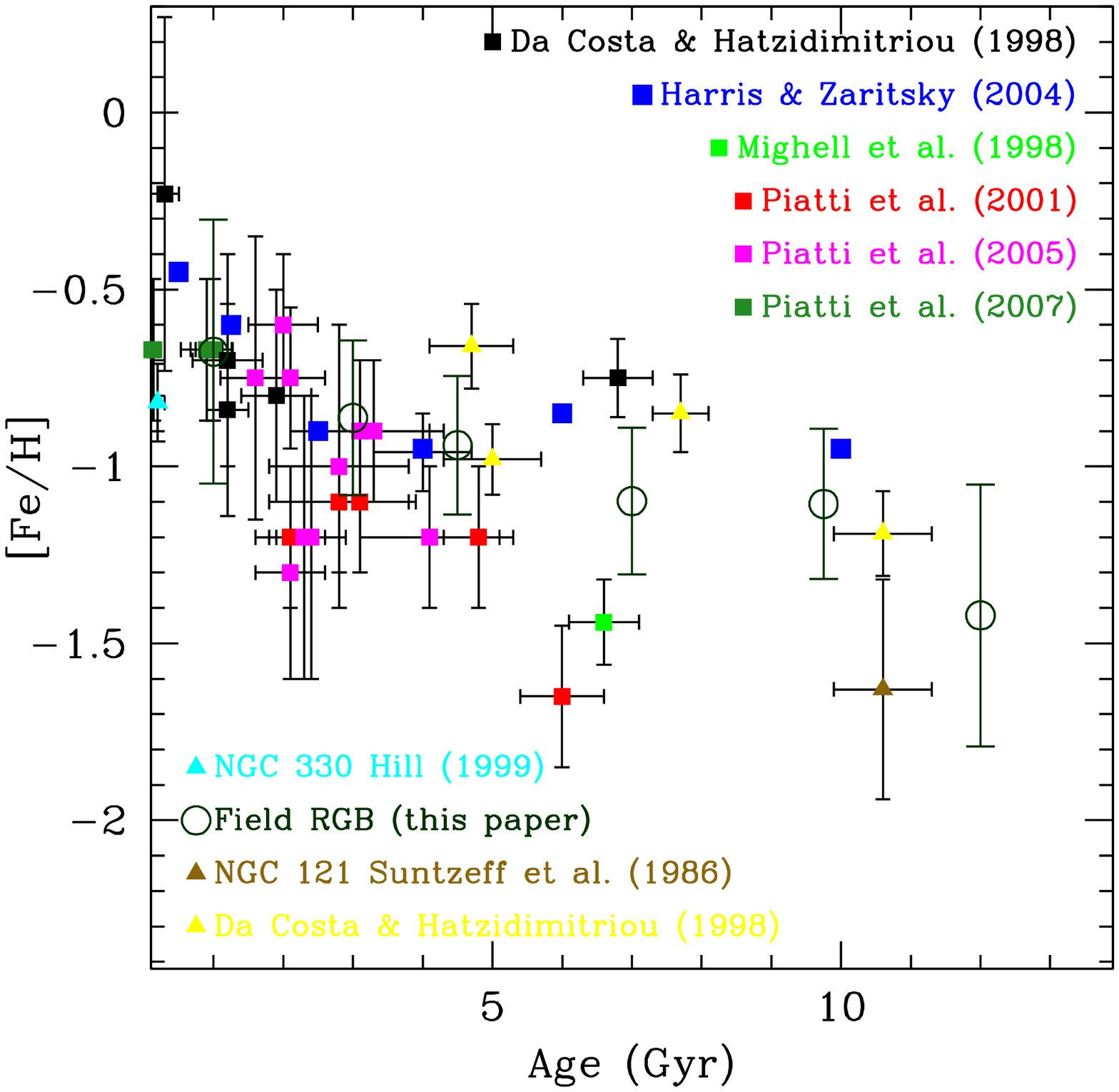}
\caption{AMR for SMC clusters. Squares represent photometric determinations: 
M98: \citet[\textsl{green}]{mighell98}; \cite[DH98, 
\textsl{black}]{costahatz98}; \citet[P01, \textsl{red}]{piatti01}; 
\citet[P05, \textsl{pink}]{piatti05a} and 
\citet[P07, \textsl{dark green}]{piatti07}. Triangles are spectroscopic 
determinations: NGC 121: \citet[S86, \textsl{brown}]{suntzeff86}; 
\citet[DH98, \textsl{yellow}]{costahatz98} and NGC 330: \citet[H99, 
\textsl{cyan}]{hill99}. Blue squares represent the AMR obtained from 
UBVI photometry by \citet[HZ04]{harriszaritsky04} in the central region of 
the galaxy. The mean metallicity in six age bins of our global SMC AMR has 
also been plotted (\textsl{open circles}). Note that the metallicity scales 
of each work may not be exactly the same.\label{cumulos}}
\end{figure}

All the AMR plotted in Figures \ref{amreste}, \ref{amroeste} and
\ref{amrsur} show a rapid chemical enrichment at a very early epoch. Even
though in some fields we have not observed enough old stars to sample this
part of the AMR, note that 12 Gyr ago all fields have reached
[Fe/H]$\simeq -1.4\sbond -1.0$. This initial chemical enrichment was followed
by a period of very slow metallicity evolution until around 3 Gyr ago.
Then, the galaxy started another period of chemical enrichment, which is
observed in the innermost fields, which are, however,the only ones where
we observed enough young stars to sample this part of the AMR. For a given
age, the mean metallicity of the stars in fields qj0111, qj0112 and
smc0057 seems to be slightly more metal-poor than those of other fields.
The uncertainties in the age determination could account for the observed
differences. In all cases, however, the mean metallicity is similar to
that of the other fields at similar galactocentric radii. The field AMRs
obtained in this work are similar to those for clusters (the reader should
take into account that differences in the metallicity scales exist among
different works), although there is only one cluster older than 10 Gyr
(see Figure \ref{cumulos}), and for planetary nebulae, with the exception
that in these objects it is not observed the chemical enrichment episode
at a very early epoch \citep[see Figure 6 of][]{idiart07}. 

 It does not seem that there was a period in which the galaxy has not
formed stars, in agreement with the result found in Paper I. (Figures
\ref{amreste}, \ref{amroeste} and \ref{amrsur}, inset panels). For eastern
fields, located in the wing, most of the observed stars have ages younger
than 8 Gyr, but there is also a significant number of objects older than
10 Gyr. At a given galactocentric distance, eastern fields show a large
number of young stars ($\leq$3 Gyr) in comparison to the western ones,
as discussed in Paper I. For the western and southern fields, the
fraction of intermediate-age stars, which are also more metal-rich,
decreases as we move away from the center, although the average
metallicity in each age bin is similar. This indicates the presence of an
age gradient in the galaxy, which may be the origin of the metallicity
one. It is noticeable that for the most external fields, qj0033 and
qj0053, we find a predominantly old and metal-poor stellar population.



We can check statistically the hypothesis that the AMR is independent of
the position in the SMC. To do so, we have combined the measurements on
the 13 fields to obtain a global AMR and compared it with that of each
field. To do so, we have divided the age range into six intervals
(age$<$1.5 Gyr, 1.5--3.5 Gyr, 3.5--5.5 Gyr, 5.5--8.5 Gyr, 8.5--11 Gyr,
$>$11 Gyr). We have computed the dispersion and mean metallicity in each
age bin, and listed the results in Table \ref{testchi2}. With these data,
we have performed a $\chi^2$ test as follows:

\begin{equation}
\chi^2=\sum_{i=1}^6\frac{(Z_i^{field}-Z_i^{global})^2}{\sigma_i^2}
 \end{equation}

where $\sigma_i^2$ is the squared sum of the uncertainties in the age bin
$i$ of each field and the global AMR. The result is shown in
Column 8 of Table \ref{testchi2}. All fields, except qj0047 and smc0033,
have values of $\chi^2_\nu$ smaller than 1. The discrepancy in fields
qj0047 and smc0033 may be due to small number statistics: there are only
two stars younger than 11 Gyr in field qj0047 and three in smc0033. Values
of $\chi^2_\nu<$1 mean that the observed AMR for each field is the same as
the global one with a confidence of 90\% (95\% in most cases). From  these
results, we may conclude that the hypothesis that the AMR is independent
of position is correct within the uncertainty.

\section{Summary and Discussion} \label{discursion}

Using CaT spectroscopy, we have derived stellar metallicities for a large
sample of RGB field stars in 13 regions of the SMC situated at different
galactocentric distances and positions angles. We have found a radial
metallicity gradient, which is most evident for those fields situated
toward the South, where we covered a large galactocentric radius. For a
given galactocentric distance, the mean metallicities for fields situated
at different position angles are very similar. The inner fields have a
mean metallicity of [Fe/H] $\sim-1$, which is similar to that of the
cluster metallicity distribution.

We have obtained the AMR of
each field from the combination of metallicities, derived from CaT
spectroscopy, and the position of stars in the CMD. All fields have
similar AMRs, which are also similar to the cluster�s one
\citep{piatti05a}. All show a rapid initial increase of metallicity,
followed by a very slow chemical enrichment period. A new relatively fast
chemical enrichment episode is observed in the last few Gyrs in the fields
within $\sim$2\arcdeg\@ of the center with enough young stars to sample
it. From the information on the AMRs, we conclude that coeval stars have
the same metallicity everywhere in the SMC. The observed metallicity
gradient is therefore related to an age gradient, because the youngest
stars, which are also the most metal-rich, are concentrated in the central
regions of the galaxy.

In a forthcoming paper we will try to reproduce the observed AMR with
chemical evolution models using accurate SFRs, as a function of time,
which are being derived by our group in each field (No\"el et al. 2008, in
preparation).

\acknowledgments

AA, CG, NEDN and RC acknowledge the support from the Spanish Ministry of
Science and Technology (Plan  Nacional de Investigaci\'on Cient\'{\i}fica,
Desarrollo, e Investigaci\'on Tecnol\'ogica, AYA2004-06343). RC also
acknowledges the funds by the Spanish Ministry of Science and Technology
under the MEC/Fullbright postdoctoral fellowship program. EC and REM
acknowledge support from the Fondo Nacional de Investigaci\'on
Cient\'{\i}fica y Tecnol\'oliga (proyecto No. 1050718, Fondecyt) and from
the Chilean Centro de Astrof\'{\i}sica FONDAP No. 15010003. This work has
made use of the IAC-STAR Synthetic CMD computation code. IAC-STAR is
supported and maintained by the computer division of the Instituto de
Astrof\'{\i}sica de Canarias.

Facilities: \facility{VLT(FORS2)}.

\clearpage


\clearpage 

\clearpage 

\clearpage 

\clearpage 

\clearpage 

\clearpage 

\clearpage 

\clearpage 

\clearpage

\clearpage






\clearpage
\begin{deluxetable}{ccccccc}
\tabletypesize{\scriptsize}
\tablecaption{SMC observed fields \label{obsfields}}
\tablewidth{0pt}
\tablehead{
\colhead{Field} & \colhead{$\alpha_{2000}$} & \colhead{$\delta_{2000}$} & \colhead{r(')} 
& \colhead{PA (\arcdeg)} & \colhead{Zone} &
\colhead{E(B-V)}
}
\startdata
{\sl smc0057} & {\sl 00:57} & {\sl -73:53} & {\sl 65.7} & {\sl 164.4} & {\sl South} & {\sl 0.09}\\
{\bf qj0037} & {\bf 00:37} & {\bf -72:18} & {\bf 78.5} & {\bf 294.0} & {\bf West} & {\bf 0.07}\\
{\bf qj0036} & {\bf 00:36} & {\bf -72:25} & {\bf 79.8} & {\bf 288.0} & {\bf West} & {\bf 0.07}\\
{\emph qj0111} & {\emph 01:11} & {\emph -72:49} & {\emph 80.9} & {\emph 89.5} & {\emph East} & {\emph 0.09}\\
{\emph qj0112} & {\emph 01:12} & {\emph -72:36} & {\emph 87.4} & {\emph 81.0 }& {\emph East} & {\emph 0.09}\\
{\bf qj0035} & {\bf 00:35} & {\bf -72:01} & {\bf 95.5} & {\bf 300.6} &{\bf  West} & {\bf 0.05}\\
{\emph qj0116} & {\emph 01:16} & {\emph -72:59} & {\emph 102.5} & {\emph 95.2} & {\emph East} & {\emph 0.08}\\
{\sl smc0100} & {\sl 01:00} & {\sl -74:57} & {\sl 130.4} & {\sl 167.5} & {\sl South} & {\sl 0.05}\\
{\sl qj0047} & {\sl 00:47} & {\sl -75:30} & {\sl 161.7} & {\sl 187.7} & {\sl South} & {\sl 0.05}\\
{\bf qj0033} & {\bf 00:33} & {\bf -70:28} & {\bf 172.9} & {\bf 325.0} & {\bf West} & {\bf 0.03}\\
{\sl smc0049} & {\sl 00:49} & {\sl -75:44} & {\sl 174.8} & {\sl 184.6} & {\sl South} & {\sl 0.06} \\
{\sl qj0102} & {\sl 01:02} & {\sl -74:46} & {\sl 179.5} & {\sl 169.4} & {\sl South} & {\sl 0.05}\\
{\sl smc0053} & {\sl 00:53} & {\sl -76:46} & {\sl 236.3} & {\sl 179.4} & {\sl South} & {\sl 0.06}\\
\enddata
\end{deluxetable}

\clearpage
\begin{deluxetable}{ccccccccc}
\tabletypesize{\scriptsize}
\tablecaption{Red giants stars observed\label{starobs}}
\tablewidth{0pt}
\tablehead{
\colhead{$\alpha_{2000}$} & \colhead{$\delta_{2000}$}& \colhead{$\Sigma$ Ca} &\colhead{$\sigma_{\Sigma Ca}$}&\colhead{B} 
& \colhead{I} & \colhead{$V_r (km s^{-1}$)} & \colhead{Comments} 
}
\startdata
00:33:53.8 & -70:24:31.3 &  2.75 &  0.23 & 20.26 & 18.45 & -307.1 &68.5 & No member\\     
00:34:00.0 & -70:28:22.9 &  2.56 &  0.42 & 20.31 & 18.57 &   79.0 & 4.2 & \\		   
00:34:03.1 & -70:27:07.7 &  1.89 &  0.71 & 20.82 & 18.97 &   78.9 & 3.3 & \\		   
00:34:25.9 & -70:27:47.1 &  5.64 &  0.19 & 19.66 & 17.61 &  133.5 & 3.7 & \\		   
00:34:06.4 & -70:27:39.3 &  1.95 &  0.12 & 18.67 & 17.81 &  162.2 & 2.9 & \\		   
00:33:37.1 & -70:27:31.8 &  5.10 &  0.08 & 18.82 & 15.82 &   69.9 & 3.1 & \\		   
00:33:41.8 & -70:27:14.3 &  5.17 &  0.20 & 19.72 & 17.46 &   31.7 & 2.9 & No member \\    
\enddata
\tablecomments{Table \ref{starobs} is published in its entirety in the electronic edition of Astronomical 
Journal. A portion is shown here for
guidance regarding its form and content}
\end{deluxetable}

\begin{deluxetable}{cc}
\tabletypesize{\scriptsize}
\tablecaption{Line and continuum bandpasses
\label{bandastable}}
\tablewidth{0pt}
\tablehead{
\colhead{Line Bandpasses (\AA)} & \colhead{Continuum bandpasses (\AA)}}
\startdata
8484-8513 & 8474-8484\\
8522-8562 & 8563-8577\\
8642-8682 & 8619-8642\\
\nodata & 8799-8725\\
\nodata & 8776-8792\\
\enddata
\end{deluxetable}

\begin{deluxetable}{ccccccc}
\tabletypesize{\scriptsize}
\tablecaption{Coefficients of the fit of equation \ref{rela} for BaSTI stellar evolution models \citep{pie04}. 
The fit standard deviation of the fit ($\sigma$) is
shown in the last column.\label{coefedad}}
\tablehead{
\colhead{a} & \colhead{b} & \colhead{c} & \colhead{d} & \colhead{e} & \colhead{f} & \colhead{$\sigma$}
}
\startdata
{\footnotesize 4.06$\pm$0.03} & {\footnotesize 3.58$\pm$0.02} & {\footnotesize
0.663$\pm$0.003} & {\footnotesize -1.314$\pm$0.006} & {\footnotesize -0.402$\pm$0.002} & {\footnotesize
-0.077$\pm$0.005} & {\footnotesize 0.40} \\
\enddata
\end{deluxetable}

\clearpage
\begin{deluxetable}{cccccc}
\tabletypesize{\scriptsize}
\tablecaption{Metallicity distribution in each field.\label{smcmetalicidades}}
\tablehead{
\colhead{Field} & \colhead{r(')} & \colhead{$\langle[Fe/H]\rangle$} & \colhead{$\sigma_{[Fe/H]}$} & \colhead{{\footnotesize
[Fe/H]$<$-1}} &
\colhead{{\footnotesize [Fe/H]$>$-1.0}}}
\startdata
{\sl smc0057} & {\sl 1.1} & {\sl -1.01} & {\sl 0.33} &  {\sl 53} & {\sl 47}\\
{\bf qj0037} & {\bf 1.3} & {\bf -0.95} & {\bf 0.17} &  {\bf 35} & {\bf 65}  \\
{\bf qj0036} & {\bf 1.3} & {\bf -0.98} & {\bf 0.25} &  {\bf 49} &
{\bf 51}\\
{\emph qj0111} & {\emph 1.3} & {\emph -1.08} & {\emph 0.21} & {\emph 64} & {\emph 36}  \\
{\emph qj0112} & {\emph 1.4} & {\emph -1.16} & {\emph 0.32} & {\emph 69} &
{\emph 31}  \\
{\bf qj0035} & {\bf 1.6} & {\bf -1.09} & {\bf 0.24} &  {\bf 66} & {\bf 34}  \\
{\emph qj0116} & {\emph 1.7} & {\emph -0.96} & {\emph 0.26} & {\emph 43} & {\emph 57} \\
{\sl smc0100} & {\sl 2.2} & {\sl -1.07} & {\sl 0.28} &  {\sl 61} & {\sl 39} \\
{\sl qj0047} & {\sl 2.7} & {\sl -1.15} & {\sl 0.27}  &  {\sl 68} & {\sl 32}  \\
{\bf qj0033} & {\bf 2.9} & {\bf -1.58} & {\bf 0.57} &  {\bf 85} & {\bf 15}  \\
{\sl smc0049} & {\sl 2.9} & {\sl -1.00} & {\sl 0.28} &  {\sl 53} & {\sl 47} \\
{\sl qj0102} & {\sl 3.0} & {\sl -1.29} & {\sl 0.42}  &  {\sl 74} & {\sl 26}  \\
{\sl smc0053} & {\sl 3.9} & {\sl -1.64} & {\sl 0.50} &   {\sl 92} & {\sl 8} \\
\enddata
\end{deluxetable}

\begin{deluxetable}{cccccccc}
\tabletypesize{\tiny}
\tablecaption{Average metallicity in six age bins. Also listed are the values obtained from the combination of
the 13 fields. The last column shows the value $\chi^2_\nu$ obtained from the comparison of each field with the
global one.\label{testchi2}}
\tablehead{
\colhead{Field} & \colhead{$\langle[Fe/H]_{\leq1.5}\rangle$} & \colhead{$\langle[Fe/H]_{1.5-3.5}\rangle$}
& \colhead{$\langle[Fe/H]_{3.5-5.5}\rangle$} & \colhead{$\langle[Fe/H]_{5.5-8.5}\rangle$} & \colhead{$\langle[Fe/H]_{8.5-11}\rangle$} & 
\colhead{$\langle[Fe/H]_{\geq11}\rangle$} & \colhead{$\chi^2$} 
}
\startdata
{\sl smc0057} & -0.68$\pm$0.11 & -0.98$\pm$0.20 & -1.09$\pm$0.14 & -1.34$\pm$0.04 & -1.24$\pm$0.05 & -1.52$\pm$0.24 & 0.39\\
{\bf qj0037} & -0.15$\pm$0.22 & -0.83$\pm$0.10 & -0.89$\pm$0.17 & -1.06$\pm$0.11 & -0.89$\pm$0.02 & -1.53$\pm$0.55 & 0.44\\
{\bf qj0036} & \nodata & -0.76$\pm$0.07 & -0.82$\pm$0.17 & -1.01$\pm$0.12 & -0.96$\pm$0.16 & -1.32$\pm$0.23 & 0.17\\
{\emph qj0111} & -0.55$\pm$0.29 & -0.96$\pm$0.19 & -1.07$\pm$0.07 & -1.17$\pm$0.13 & -1.56$\pm$0.04 & -1.98$\pm$0.22 & 0.62\\
{\emph qj0112} & -0.85$\pm$0.34 & -1.03$\pm$0.28 & -1.23$\pm$0.06 & -1.38$\pm$0.20 & \nodata & -1.77$\pm$0.27 & 0.76\\
{\bf qj0035} & \nodata & -0.66$\pm$0.24 & -0.73$\pm$0.26 & -0.98$\pm$0.19 & -0.94$\pm$0.14 & -1.27$\pm$0.26 & 0.47\\
{\emph qj0116} & -0.71$\pm$0.68 & -0.56$\pm$0.12 & -0.90$\pm$0.15 & -1.06$\pm$0.20 & -1.03$\pm$0.07 & -1.22$\pm$0.26 & 0.30\\
{\sl smc0100} & \nodata & -0.80$\pm$0.20 & -0.91$\pm$0.16 & -0.89$\pm$0.02 & -1.16$\pm$0.06 & -1.30$\pm$0.26 &  0.24\\
{\sl qj0047} & \nodata &   -0.63$\pm$0.01 & -0.67$\pm$0.01 & \nodata & \nodata & -1.20$\pm$0.38 & 1.07\\
{\bf qj0033} & \nodata & \nodata & \nodata & -0.74$\pm$0.01 & -1.05$\pm$0.18 & -1.75$\pm$0.40 & 1.07\\
{\sl smc0049} & \nodata & -0.74$\pm$0.14 & -0.87$\pm$0.06 & -0.92$\pm$0.02 & -1.13$\pm$0.04 & -1.28$\pm$0.17 & 0.24\\
{\sl qj0102} & \nodata & -1.00$\pm$0.28 & -0.99$\pm$0.18 & -1.06$\pm$0.38 & -1.31$\pm$0.15 & -1.49$\pm$0.31 & 0.17\\
{\sl smc0053} & \nodata & \nodata & \nodata & \nodata & -1.27$\pm$0.05 & -1.67$\pm$0.43 & 0.36\\
Global & -0.68$\pm$0.37 & -0.86$\pm$0.22 & -0.94$\pm$0.20 & -1.10$\pm$0.21 & -1.11$\pm$0.21 & -1.42$\pm$0.37 &\nodata \\
\enddata
\end{deluxetable}

\end{document}